\documentclass{aa}

\usepackage{txfonts}
\usepackage{natbib}
\usepackage{units}
\usepackage{graphicx}
\usepackage{url}
\bibpunct{(}{)}{;}{a}{}{,}

\newcommand{\be}{\begin{equation}}
\newcommand{\ee}{\end{equation}}
\newcommand{\nel}{\ensuremath{n_\mathrm{e}}}
\newcommand{\bifrost}{{\textsl{Bifrost}}}
\newcommand{\HI}{\ion{H}{i}}

\newcommand{\CaII}{\ion{Ca}{ii}}
\newcommand{\CaIII}{\ion{Ca}{iii}}
\newcommand{\MgII}{\ion{Mg}{ii}}
\newcommand{\hk}{h\,\&\,k}

\newcommand{\eg}{{\it e.g.,}} 
\newcommand{\dd}{\, \mathrm{d}}

\title{Approximations for radiative cooling and heating  in the
  solar chromosphere}

\author{M. Carlsson\inst{1,2}
  \and  J. Leenaarts\inst{1,3}
}

\institute{
Institute of
Theoretical Astrophysics, University of Oslo, P.O. Box 1029
Blindern, N-0315 Oslo, Norway
\and
Center of Mathematics for Applications,
University of Oslo, P.O. Box 1053
Blindern, N-0316 Oslo, Norway
\and
Sterrekundig Instituut, Utrecht University, Postbus 80\,000,
NL--3508 TA Utrecht, The Netherlands
}

\date{\today /- }

\authorrunning{Carlsson \& Leenaarts}
\titlerunning{Radiative energy balance recipes}
\abstract
{The radiative energy balance in the solar chromosphere is dominated by strong spectral lines that are formed out of LTE. It is computationally prohibitive to solve the full equations of radiative transfer and statistical equilibrium in 3D time dependent MHD simulations.}
{To find simple  recipes  to compute the radiative energy balance in the dominant lines 
under solar chromospheric conditions.}
{We use detailed calculations in time-dependent and 2D MHD snapshots to derive empirical
formulae for the radiative cooling and heating.}
{The radiative cooling in neutral hydrogen lines and the Lyman
  continuum, the H and K and intrared triplet lines of singly ionized
  calcium and the h and k lines of singly ionized magnesium can be
  written as a product of an optically thin emission (dependent on
  temperature), an escape probability (dependent on column mass) and an ionization fraction (dependent on temperature). In the cool pockets of the chromosphere the same transitions contribute to the heating of the gas and similar formulae can be derived for these processes. We finally derive a simple recipe for the radiative heating of the chromosphere from incoming coronal radiation. We compare our recipes with the detailed results and comment on the accuracy and applicability of the recipes.}
{} 
\begin{document}

\maketitle

\section{Introduction}\label{sec:introduction}

The chromospheric radiative energy balance is dominated by a small
number of strong lines from neutral hydrogen, singly ionized calcium
and singly ionized magnesium
\citep{Vernazza+Avrett+Loeser1981}
although a large number of iron lines may be equally important in the
lower atmosphere
\citep{Anderson+Athay1989}. The strong lines are formed out of Local
Thermodynamic Equilibrium (LTE) which means that numerical simulations
of the dynamic chromosphere need to solve simultaneously the
(magneto-) hydrodynamic equations, radiative transfer equations and
rate equations for all radiative transitions and energy
levels involved. This is a major computational effort that has been
accomplished in one dimensional simulations of wave propagation in the
quiet solar atmosphere
\citep{Klein+Kalkofen+Stein1976,
 Klein+Stein+Kalkofen1978,
 Carlsson+Stein1992,
 Carlsson+Stein1994,
 Carlsson+Stein1995,
 Carlsson+Stein1997,
 Carlsson+Stein2002,
 Rammacher+Ulmschneider2003,
 Fossum+Carlsson2005,
 Fossum+Carlsson2006}, 
sunspots \citep{Bard+Carlsson2010}, 
solar flares \citep{Abbett+Hawley1999,
  Allred+Hawley+Abbett+Carlsson2005, 
  Kasparova+Varady+Heinzel+Karlicky+etal2009,
  Cheng+Ding+Carlsson2010}
and stellar flares \citep{Allred+Hawley+Abbett+Carlsson2006}. 

In three-dimensional simulations the computational work required to solve
the full set of equations becomes prohibitive; there is a need for a
simplified description that retains as much as possible of the basic
physics but brings the computational work to tractable levels. The aim
of this paper is to provide such simplified recipes for the
chromospheric radiative energy balance.

In addition to the cooling in strong lines and continua the same
transitions may provide heating of cool pockets in the
chromosphere. We also get heating in the chromosphere from coronal
radiation impinging from above. We will here address all three
processes: cooling from lines and continua, heating from the same
transitions and heating through absorption of coronal radiation.

We use detailed radiative transfer calculations from dynamical
snapshots to derive simple lookup tables that only use easily
obtainable quantities from a single snapshot from a simulation. We
test these recipes by applying them to several simulation runs and
compare the radiative losses from detailed solutions with those from
the recipes and find in general good agreement.

The structure of the paper is as follows: 
In Sec.~\ref{sec:atmospheres} we describe the hydrodynamical snapshots 
used for deriving the lookup tables, 
in Sec.~\ref{sec:atoms} we describe the atomic models used, 
in Sec.~\ref{sec:cooling} we derive the recipes for radiative cooling from
strong lines of hydrogen, calcium and magnesium, 
in Sec.~\ref{sec:heating} we derive the recipes for heating of cool pockets of gas
from the same lines, 
in Sec.~\ref{sec:comparison} we compare the recipes with a detailed calculation,
in Sec.~\ref{sec:incrad} we derive the recipes for the heating by incident coronal radiation
and in Sec.~\ref{sec:conclusions} we end with our conclusions.

\section{Atmospheric models}\label{sec:atmospheres}

To calculate the semi-empirical recipes we use two sets of
hydrodynamical snapshots. The first is a simulation of wave
propagation in the solar atmosphere using the RADYN code (Carlsson \&
Stein 1992, 1994, 1995, 1997, 2002). RADYN solves the one-dimensional
equations of conservation of mass, momentum, energy and charge
together with the non-LTE radiative transfer and population rate
equations, implicitly on an adaptive mesh.  The grid positions of the
adaptive mesh are determined by a grid equation that specifies where
high resolution is needed (typically where there are large
gradients). The grid equation is solved simultaneously with the other
equations. The simulation includes a 5-level plus continuum hydrogen
atom, 5-levels of singly ionized calcium and one level of doubly
ionized calcium and a 9-level helium atom including 5 levels of
neutral helium, 3 of singly ionized helium and one level of doubly
ionized helium. All allowed radiative transitions between these levels
are included. At the bottom boundary we introduce a velocity field
that was determined from MDI observations of the Doppler shifts of the
676.78 nm \ion{Ni}{i} line. Coronal irradiation of the chromosphere
was included using the prescription of 
\citet{Wahlstrom+Carlsson1994}
based on 
\citet{Tobiska1991}.
The effects of non-equilibrium ionization
are handled self-consistently.  The simulation is essentially the one
used in \citet{Judge+Carlsson+Stein2003} but with updated atomic data
for calcium. A number of  ingredients
that are potentially important for the chromospheric energy balance
are missing: line blanketing is not included (especially the iron
lines may be important \citep{Anderson+Athay1989}),  cooling in the strong
\ion{Mg}{ii} $h$ and $k$ lines is not
included and the description of partial redistribution is
very crude for hydrogen and altogether missing for calcium. Most
important is probably the restriction of only one dimension. This
prevents shocks from spreading out, forces shock-merging of overtaking
shocks and prevents the very low temperatures found in 2D and 3D
simulations from strong expansion in more than one dimension. 

The other snapshot is from a 2D simulation of the solar atmosphere
spanning from the upper convection zone to the corona. This simulation
was performed with the radiation-magneto-hydrodynamics code \bifrost\
\citep{BIFROST} and includes the effects of non-equilibrium ionization
of hydrogen using the methods described in
\citet{Leenaarts+Carlsson+Hansteen+Rutten2007}.
A detailed description of the setup and results of this simulation is
given in
\citet{Leenaarts+Carlsson+Hansteen+Gudiksen2011}.
Also here we have shortcomings in the description. Line blanketing is
included but with a two-level formulation for the scattering probability,
hydrogen non-equilibrium ionization is included but the excitation
balance (needed to calculate the individual lines) is very crude.

Both simulations thus have their shortcomings. The purpose of this
paper is, however, to develop recipes that reproduce the results of
detailed non-LTE calculations. It is thus not crucial that the
simulations reproduce solar observations in detail as long as they
cover the parameter space expected in the solar chromosphere. 

For hydrogen it
is important to include the coupling between the non-equilibrium
ionization and the non-LTE excitation and we therefore use the RADYN
simulation for the hydrogen recipes.
For magnesium and calcium it is more important to include PRD and have
a large temperature range than
to include effects of non-equilibrium ionization and we choose the
\bifrost\ 2D simulations. Non-equilibrium ionization is not important 
for calcium \citep{Wedemeyer-Bohm+Carlsson2011} while 
\citet{Rammacher+Ulmschneider2003} and test-calculations with RADYN
indicate that ionization/recombination times are as long as 100-500~s 
for magnesium and therefore of importance for the ionization balance.
However, we will see that  magnesium is almost exclusively in the form of
\MgII\ in the temperature range where magnesium is important for the
cooling/heating of the chromosphere and the neglect of non-equilibrium
ionization is therefore not important for the results.

\section{Atomic models}\label{sec:atoms}

\subsection{Hydrogen}\label{sec:hydrogen}

We take the population densities directly from the RADYN simulation
and thus use the same atomic parameters.
We use a five-level plus continuum model atom with energies from
\citet{Bashkin+Stoner1975}, oscillator strengths from 
\citet{Johnson1972}, photoionization cross-sections from 
\citet{Menzel+Pekeris1935} and
collisional excitation and ionization rates from
\citet{Vriens+Smeets1980} except for Lyman-$\alpha$ where we use
\citet{Janev+Langer+Evans1987}.
We use Voigt profiles with complete redistribution (CRD) for all
lines. For the Lyman lines, profiles truncated at six Doppler widths
are used to mimic effects of partial redistribution (PRD)
\citep{Milkey+Mihalas1973}.

\subsection{Calcium}\label{sec:calcium}

\begin{figure}
 \includegraphics{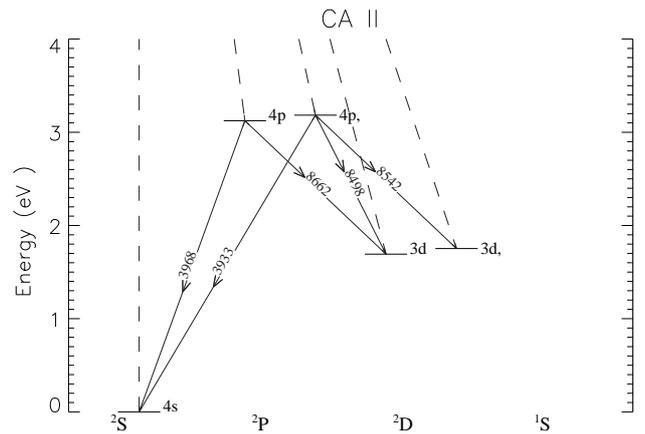}
 \caption{\label{fig:caterm} Term diagram of the \CaII\ model atom. The
 continuum is not shown. Energy levels (horizontal
 lines), bound-bound transitions ({\it solid}) and bound-free transitions
 ({\it dashed}). }
\end{figure}

We use a five-level plus continuum model atom. It contains the five
lowest energy levels of \CaII\ and the \CaIII\ continuum (see
Fig.~\ref{fig:caterm}). The energies are from the NIST database with
data coming from \citet{Sugar+Corliss1985}, the oscillator strengths
are from \citet{Theodosiou1989} with lifetimes agreeing well with the
experiments of \citet{Jin+Church1993} and the broadening parameters
are from the Vienna Atomic Line Database (VALD)
\citep{Piskunov+Kupka+Ryabchikova+Weiss+etal1995,Kupka+Piskunov+Ryabchikova+Stempels+etal1999}.
Photoionization cross-sections are from the Opacity project.
Collisional excitation rates are from
\citet{Melendez+Bautista+Badnell2007} extrapolated beyond the
tabulated temperature range of 3000~K to 38000~K using
\citet{Burgess+Tully1992}. Collisional ionization rates from the
ground state are from \citet{Arnaud+Rothenflug1985} and from excited states
we use the general formula provided by
\citet{Burgess+Chidichimo1983}. Autoionization is treated according to 
\citet{Arnaud+Rothenflug1985} and dielectronic recombination according
to \citet{Shull+van-Steenberg1982}. The abundance is taken from 
\citet{Asplund+Grevesse+Sauval+Scott2009}.
We assumed partial frequency redistribution (PRD) of radiation over
the line profiles for the H\&K lines.

\subsection{Magnesium}\label{sec:magnesium}

We use a three-level plus continuum model atom, including the ground
state, the upper levels of the \hk\ lines and the continuum.
Energy levels and oscillator strengths are from the NIST database,
collisional excitation rates are from \citet{Sigut+Pradhan1995}, 
collisional ionization from Seaton's semi-empirical formula
\citep[][page 42]{Allen1964}
and photoionization cross-sections are
from the Opacity Project\footnote{\url{http://cdsweb.u-strasbg.fr/topbase/TheOP.html}}. The abundance is taken from 
\citet{Asplund+Grevesse+Sauval+Scott2009}.
The \hk\ lines are computed assuming PRD.

\section{Cooling from lines and continua}\label{sec:cooling}

The basic cooling process is an electron impact excitation followed by a
radiative deexcitation. If this photon escapes the atmosphere, the
corresponding energy has been taken out of the thermal pool which
corresponds to radiative cooling. In the corona these processes are
the dominant ones: collisional excitation followed by radiative
deexcitation with the photon escaping. Collisional de-excitation and
absorption of photons (radiative excitation) are insignificant in
comparison. This is often called the coronal approximation. 

In the
chromosphere we have a much more complex situation with all rates
playing a role. In strong lines most of the photons emitted are
immediately absorbed in the same transition but the collisional
destruction probability is often very low and the photon may escape
after a very large number of scattering events. We may also have
photons absorbed in one line (\eg\ Lyman-$\beta$) escaping in another
transition (\eg\ H-$\alpha$). A full description of these processes
entails solving the transfer equations coupled with the non-LTE rate
equations (or statistical equilibrium equations if rates are assumed
to be instantaneous). 

We here choose to describe the net effect as a
combination of three factors that need to be determined empirically
from our simulations: {\it (1)} the optically thin radiative loss function (that
gives the energy loss from the total generation of local photons in
the absence of absorption per atom in the ionization stage of
consideration (\HI, \CaII\ and \MgII)), {\it (2)} the probability that this
energy escapes from the atmosphere and {\it (3)} the fraction of atoms in the
given ionization stage:

\be
Q_{\rm X} = - L_{{\rm X}_m}(T)E_{{\rm X}_m}(\tau){N_{{\rm X}_m}\over N_{\rm
    X} }(T) A_{\rm X}{N_{\rm H}\over\rho}\nel\rho \label{eq:Q}
\ee

\noindent where $Q_{\rm X}$ is the radiative heating (positive $Q$) or cooling
(negative $Q$) per volume from element X (in our case H, Ca and Mg), 
$L_{{\rm X}_m} (T)$ is the optically thin radiative loss
function as function of
temperature, $T$, per electron, per particle of element X in
ionization stage $m$, $E_{{\rm X}_m}(\tau)$ is the escape probability
as function of some depth parameter $\tau$, ${N_{{\rm X}_m}\over N_{\rm
    X} }(T)$ is the fraction of element X that is in ionization
stage $m$, $A_{\rm X}$ is the abundance of element X, ${N_{\rm
    H}\over\rho}$ is the number of hydrogen particles per gram of
stellar material, which is a constant for a given set of abundances
(=4.407$\times 10^{23}$ for the solar abundances of
\citet{Asplund+Grevesse+Sauval+Scott2009}), $\nel$ is the electron number density
and $\rho$ is the mass density.

The optically thin radiative loss function is equal to the 
collisional excitation rate in the coronal approximation and is then only a
function of temperature. In the chromosphere this is not the case but
the hope is that we may still write a radiative loss function as
function of temperature only. We expect various routes between the
levels (rather than a collisional excitation in one transition
followed by a radiative deexcitation in the same transition) but as
long as collisional de-excitation rates are small we expect the
{\em total} collisional excitation from the ground state to be equal
to the radiative losses:

Assume a gas where 
\be
R_{ji} \gg \sum_{k=1}^{n_\mathrm{l}}C_{jk}, i<j,  \label{eq:rgtc}
\ee
\be
\sum_{k=1}^{j-1} R_{jk} \gg \sum_{k=j+1}^{n_\mathrm{l}}R_{jk}, \label{eq:rgtr}
\ee
Here $R_{ji}$ is the radiative rate coefficient from bound level $j$
to $i$ and $C_{jk}$ the collisional rate coefficient from level $j$ to
$k$ and $n_\mathrm{l}$ is the number of energy levels of the atom. 
These assumptions imply that an excited atom will nearly always
fall back to its ground state through one or more radiative
transitions. Hence, photon absorption will always be followed by one
or more radiative deexcitations, and the gas cannot absorb energy from
the radiation field. The only way to lose energy from the gas through
a bound-bound transition is through a collisional excitation followed
by one or more radiative deexcitations. All energy of the initial
excitation is then emitted as radiation and thermal energy is
converted into radiation.

Therefore, as long as Eqs.~\ref{eq:rgtc}--\ref{eq:rgtr} hold, the
radiative losses $L_{\rm X_m}$ due to lines of atoms of species X in
ionization stage $m$, per atom in ionization stage $m$ per electron, is
given by
\be
L_{\rm X_m} = \frac{N_{{\rm X}_{m,1}}}{N_{{\rm X}_m}}\sum_{j=2}^{n_\mathrm{l}} \frac{ \chi_{1j}C_{1j}}{\nel}, \label{eq:exc}
\ee
where $N_{{\rm X}_{m,1}}$ is the population density in the ground state
of element X, ionization stage $m$, $\chi_{1j}$ is the energy
difference between the ground state (level 1) and  level
$j$. Collisions with electrons are dominant under typical
chromospheric conditions and $\frac{C_{1j}}{\nel}$ and therefore also
$L_{\rm X_m} $ will be a function of temperature only (apart from the
factor $\frac{N_{{\rm X}_{m,1}}}{N_{{\rm X}_m}}$ which is close to
one). There are therefore reasons to believe that we may write the
optically thin loss function as a function of temperature only. In
Sec.~\ref{sec:excitation} we will use our numerical simulations to
deduce such relations and compare them with the total collisional
excitation from the ground state (Eq.~\ref{eq:exc}).

In the chromosphere not all photons will escape the atmosphere. We
parametrize this with the escape probability function $E_{{\rm
  X}_m}$ in Eq.~\ref{eq:Q}. In Sec.~\ref{sec:escape} we use our simulations
to calculate this escape probability empirically. 

We complete the recipe by determining in Sec.\ref{sec:ionization} the
factor ${N_{{\rm X}_m}\over N_{\rm X} }(T)$ in Eq.~\ref{eq:Q}: the
fraction of hydrogen, calcium and magnesium that resides in \HI,
\CaII\ and \MgII.

\subsection{Optically thin radiative loss function}\label{sec:excitation}

From the RADYN simulation we calculate the total radiative losses as
the sum of the net downward radiative rates multiplied with the energy
difference of the transition, summed over all bound-bound transitions
and the Lyman-continuum transition (the other continuum transitions
get thin low down in the atmosphere and their influence on the energy
balance can be taken into account in a multi-group opacity approach).

\begin{figure}
 \includegraphics{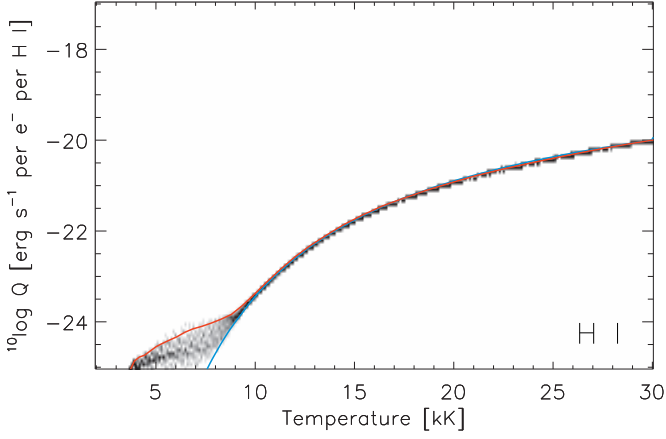}
 \caption{\label{fig:exc_h} Probability density function of
   the radiative losses in lines and the Lyman-continuum of \HI\ in the RADYN test atmosphere
   as function of temperature for points above 1.7 Mm height.
    The curves give the adopted fit ({\it red}) and the total collisional excitation according to 
    Eq.~\ref{eq:exc} ({\it blue}).}
\end{figure}

Figure~\ref{fig:exc_h} shows the probability density
function (PDF) of the total radiative losses per neutral hydrogen atom
per electron as function of temperature for the RADYN simulation\footnote{This figure, and many of the following figures, are not scatterplots
but 2D histograms normalized for each x-axis bin. The darkness at a
y-axis bin is thus a measure of the probability of finding an atmospheric
point at this particular y-value given that it is in the x-axis bin.}.
We exclude the startup phase of
the first 1000~s of the simulation and only include points above
1.7~Mm height to avoid most points where the escape probability is
small.  For
temperatures above 10 kK the relation is very tight and the PDF is
close to the total excitation curve. Below 10 kK we have a larger
spread, both because this part includes points with a low escape
probability and because the approximations (Eqns. \ref{eq:rgtc}--\ref{eq:rgtr})
start to break down. At low temperatures, the optical depth in the Lyman transitions
builds up quickly when integrating downwards in the atmosphere and only the uppermost of 
these points have an escape probability close to one.
To
account for points with low escape probability we therefore take the upper
envelope as the adopted optically thin radiative loss function.

\begin{figure}
 \includegraphics{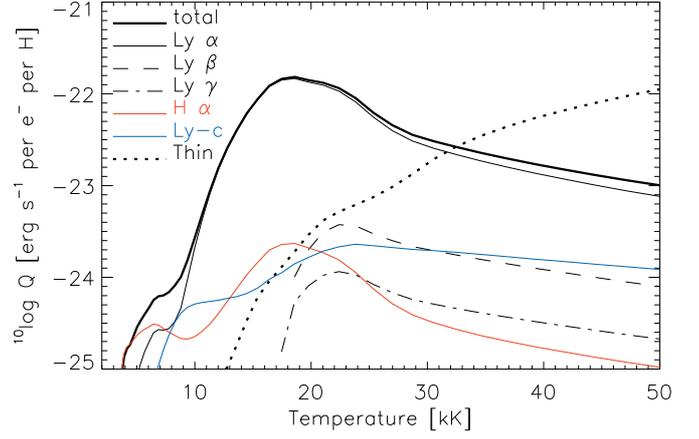}
 \caption{\label{fig:hcool_kr} Fit to the optically thin cooling per
   electron per hydrogen particle ({\it thick solid}), contributions
   from Lyman-$\alpha$ ({\it thin black solid}), Lyman-$\beta$ ({\it
     dashed}), Lyman-$\gamma$ ({\it dot-dashed}), H-$\alpha$ ({\it
     red}), Lyman-continuum ({\it blue}) and optically thin radiative
   losses from other elements than hydrogen ({\it thick dotted}) as
   function of temperature.}
\end{figure}

Figure~\ref{fig:hcool_kr} shows the various contributions to the
radiative cooling in hydrogen, this time per electron per hydrogen
atom (and not per neutral hydrogen atom) to have a normalization
similar to the one used for coronal optically thin radiative
losses. In most of the temperature range, the Lyman-$\alpha$
transition dominates with at most 20\% coming from other transitions
(where Lyman-$\beta$ and the Lyman-continuum are the most important
ones). At the low temperature end, the Lyman-continuum starts to
contribute and H-$\alpha$ eventually dominates below 7 kK. The
hydrogen optically thin radiation dominates over contributions from
other elements below 32 kK.

\begin{figure}
 \includegraphics{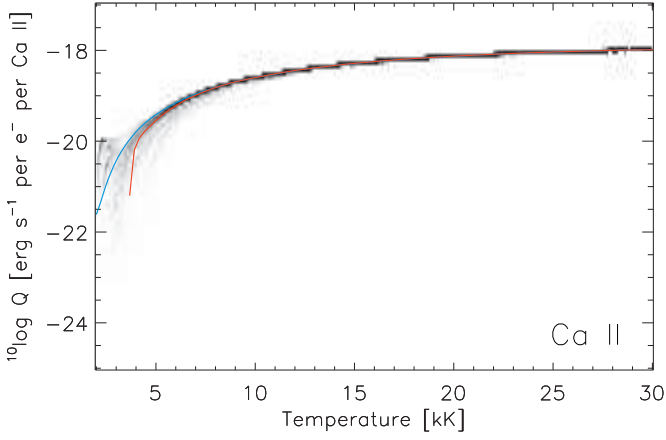}
 \caption{\label{fig:exc_ca} Probability density function of
   the radiative losses in lines  of \CaII\ as function of temperature
   in the \bifrost\ test atmosphere for points above 1.3 Mm height. 
    The curves give the adopted fit ({\it red}) and the total collisional excitation according to 
    Eq.~\ref{eq:exc} ({\it blue}).}
\end{figure}

\begin{figure}
 \includegraphics{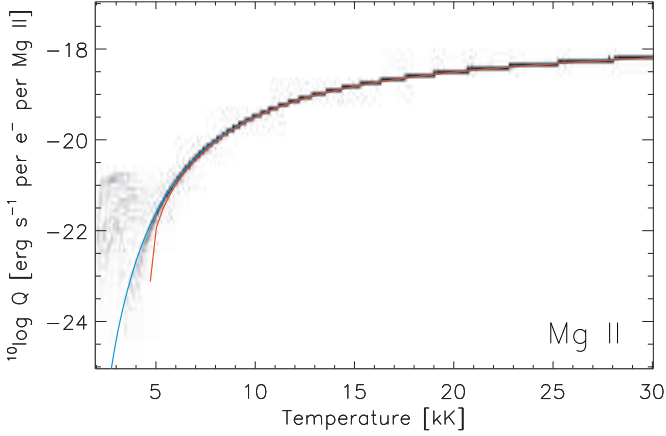}
 \caption{\label{fig:exc_mg} Probability density function of
   the radiative losses in lines  of \MgII\ as function of temperature
   in the \bifrost\ test atmosphere for points above 1.3 Mm height.
    The curves give the adopted fit ({\it red}) and the total collisional excitation according to 
    Eq.~\ref{eq:exc} ({\it blue}).}
\end{figure}

Figures~\ref{fig:exc_ca}--\ref{fig:exc_mg} show the
probability density function of the total radiative losses per
singly ionized calcium and magnesium atom, respectively, as function of temperature
for the \bifrost\ simulation.  This cooling was computed in detail using the
radiative transfer code Multi3d
\citep{Leenaarts+Carlsson2009}
assuming statistical equilibrium and plane-parallel geometry with each
column in the snapshot treated as an independent 1D atmosphere.

The probability density function is close to the total collisional
excitation curve for both \CaII\ and \MgII. In the case of \CaII\ this
may be a bit surprising at first glance.  The \CaII\ 3d\,$^{2}$D
doublet is metastable and we do not include transitions from it to the
ground state. This implies Eq.~\ref{eq:rgtc} does not hold. However,
for typical chromospheric electron densities, gas temperatures and
radiation temperatures in the infrared triplet the following relation
holds:
\be
R_{ij} \gg \sum_{k=1}^{n_\mathrm{l}}C_{ik}, j > i.  \label{eq:rgtc2}
\ee
This means that an atom in a \CaII\ 3d\,$^{2}$D state is much more likely to
get radiatively excited to one of the \CaII\ 4p\,$^{2}$P states than to move to
another level through collisions. Therefore, transition chains
starting with collisional excitation from the ground state to a
metastable level typically go through a number of radiative transitions,
and ultimately end in the ground state through a
spontaneous emission from a \CaII\ 4p\,$^{2}$P state. The original collision
energy is then lost from the gas and Eq.~\ref{eq:exc} still holds.
However, net collisional rates between the excited levels give 
a larger spread and also radiative losses that sometimes are {\em smaller} than the
total collisional excitation from the ground state. In contrast to the case of hydrogen,
the spread is thus not caused by optical depth effects and hence we take the average
of the values for our fits rather than the upper envelope. At the lowest temperatures we
also get points with radiative heating (not visible in the logarithmic plots) and the fits
make a sharp dip downwards. See 
Sec.~\ref{sec:heating} for a discussion of radiative heating.

\subsection{Escape probability}\label{sec:escape}

The assumption that all collisional excitations lead to emitted
photons breaks down in the chromosphere, where increased mass density
leads to increased probability of absorption of photons and
collisional deexcitation.  Once radiative excitation processes come
into play we have many more possible routes in the statistical
equilibrium. To fully account for these processes we have to solve the
coupled radiative transfer equations and non-LTE rate
equations. Instead, we calculate an empirical escape probability
from the ratio of the actual cooling as obtained from a detailed
radiative transfer computation and the optically thin cooling function
determined in Sec.~\ref{sec:excitation}. Ideally we would like to
tabulate this escape probability as a function of optical depth in the
lines. Since we are dealing with the total cooling over a number of
lines, there is no single optical depth scale to be chosen. 
For \CaII\ and \MgII\ we choose to tabulate the escape probability as
a function of column mass. The rationale behind this choice is that
the optical depth in the resonance lines is proportional to column
mass as long as the ionization stage under consideration is the
majority species. This is the case for both \CaII\ and \MgII, as shown
in Sec.~\ref{sec:ionization}, in the region of the atmosphere where the
optical depth is significant. Another advantage with the choice of
column mass is that it is easy to calculate from the fundamental
variables in a snapshot of an MHD simulation. For hydrogen the situation is different. The
Lyman lines get optically thin in the lower part of the transition
region where hydrogen gets ionized. In the 1D simulation used for
calculating the empirical coefficients, the transition region stays at
a constant column mass. On a column mass scale, the hydrogen escape
probability would then be a very steep function. Instead, we need to
use a depth-variable that is proportional to the neutral hydrogen
column density. We use the total column density of neutral
hydrogen multiplied with the constant 4$\times 10^{-14}$~cm$^2$ which
gives a depth-variable that is close to the optical depth at line
center of Lyman-$\alpha$. The escape probability is high to rather
high optical depths, because of the very large scattering
probability in the Lyman transitions. 

The escape probability depends in a complicated way on the density,
electron density, temperature and local radiation field. We expect it
to be close to unity at low column mass density, and to go to zero at
large column mass density.

In order to derive an empirical relation we calculate the average (over time 
for the RADYN simulation and over space for the \bifrost\ snapshot) actual
cooling and the average cooling according to the optically thin cooling function
determined in Sec.~\ref{sec:excitation} as function of the chosen depth-scale.
We adopt this ratio as the empirical escape probability as function of depth.

\begin{figure}
 \includegraphics{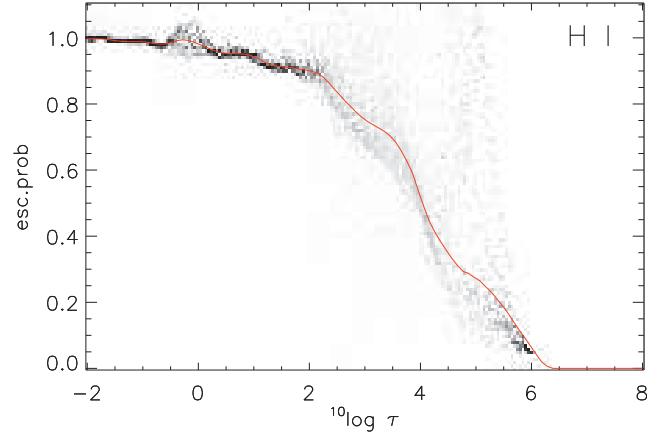}
 \caption{\label{fig:esc_h} Probability density
   function of  the empirical escape probability as
   function of approximate optical depth at Lyman-$\alpha$ line center and the
   adopted fit ({\it red}). Only points with a cooling above 
   5$\times 10^9$  erg~s$^{-1}$~g$^{-1}$ are
   included and the startup phase of the simulation (first 1000~s)
   has been excluded. }
\end{figure}

\begin{figure}
 \includegraphics{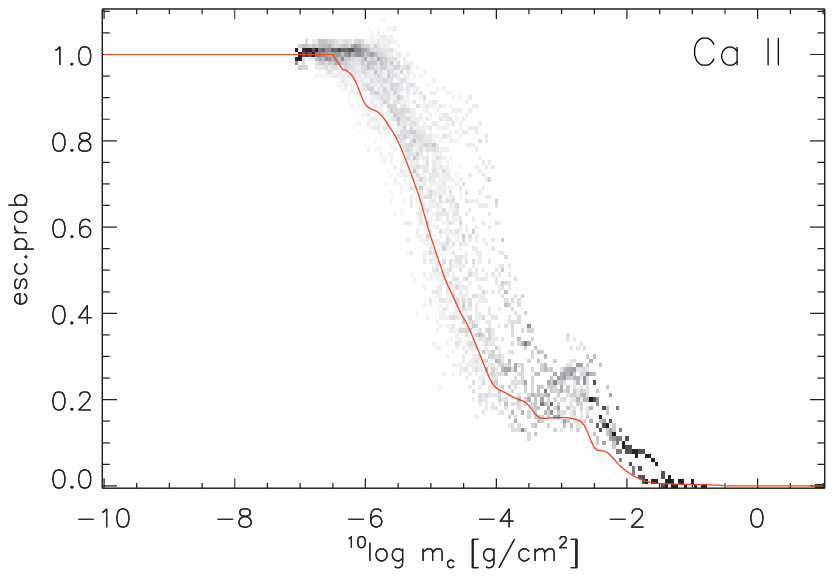}
 \caption{\label{fig:esc_ca} Probability density
   function of  the empirical escape probability as
   function of column mass for \CaII\ and the adopted fit ({\it red}). Only
   points with a a cooling above
   5$\times 10^8$  erg~s$^{-1}$~g$^{-1}$ are included.}
\end{figure}

\begin{figure}
 \includegraphics{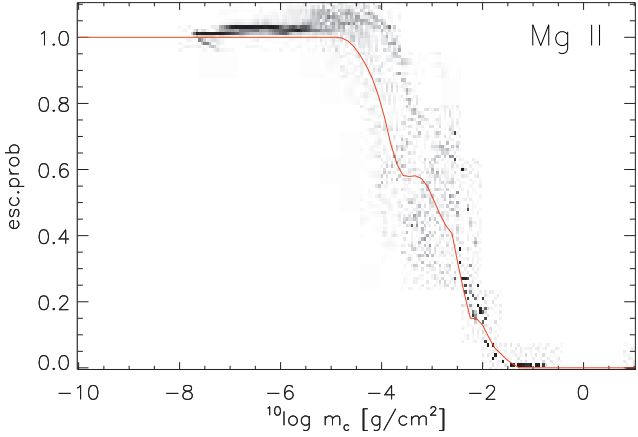}
 \caption{\label{fig:esc_mg} Probability density
   function of  the empirical escape probability as
   function of column mass for \MgII\ and the adopted fit ({\it red}). Only
   points with a a cooling above
   5$\times 10^8$  erg~s$^{-1}$~g$^{-1}$ are included.}
\end{figure}

Figure~\ref{fig:esc_h} shows the probability density function of
the empirical escape probability as function of approximate optical depth at
Lyman-$\alpha$ line center together with the adopted relation.
Figs.~\ref{fig:esc_ca} and~\ref{fig:esc_mg} show the
probability density function of the empirical escape probability as
function of column mass for \CaII\ and \MgII\ together with the adopted 
relation.

Note that the adopted relations are not the averages of the points shown in 
Figs.~\ref{fig:esc_h}--\ref{fig:esc_mg} but the ratio between the 
average of the actual cooling and the average of the optically thin cooling such
that points with large cooling get larger weight. The adopted fit may therefore
look like a poor representation of the PDFs of Figs.~\ref{fig:esc_h}--\ref{fig:esc_mg}.
If these figures had only included the points with the largest cooling, the correspondence
would have been more obvious.

\subsection{Ionization fraction}\label{sec:ionization}

In Sections.~\ref{sec:excitation}--\ref{sec:escape} we derived expressions for the radiative
losses per atom in the relevant ionization stage per electron. In order to convert
these losses to actual losses per volume, the fraction of atoms in the
ionization stage under study (\HI, \CaII\ and \MgII) must be known. In
general this quantity is a function of the electron density,
temperature, radiation field, and, in case of small transition
rates, the history of the atmosphere.

Neither Saha ionization equilibrium nor coronal equilibrium are valid
in the chromosphere. Therefore no quick physics-based recipe exists
that gives the ionization fraction. Fortunately, in our 1D and 2D test atmospheres, the
ionization fraction as function of temperature shows a rather small spread around an
average value and an empirical fit to these values gives a reasonable
approximation. We also computed fits as function of temperature and
electron density. As the ionization fraction is relatively insensitive to the electron
density these two-parameter fits do not yield a significantly better
approximation and we settle for relations as function of temperature only.

\begin{figure}
 \includegraphics{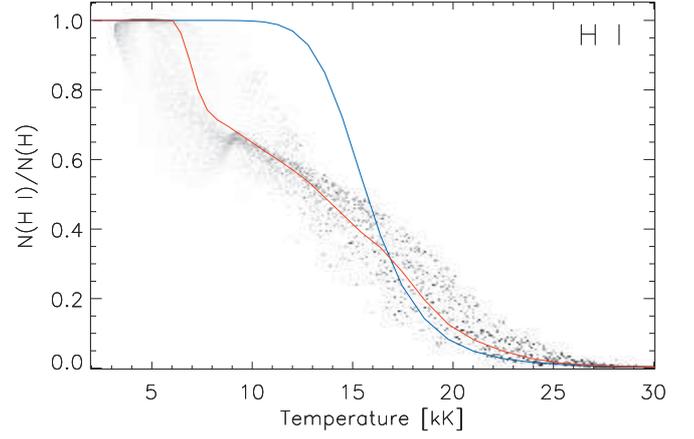}
 \caption{\label{fig:ion_h_radyn}  Probability density function of
   the fraction of neutral H atoms as function of temperature in the RADYN test
   atmosphere between heights of 0.5~Mm and 2.0~Mm after the initial
   1000~s. 
   Curves show the adopted fit to the PDF ({\it red}), and 
   the coronal approximation ({\it blue}). }
\end{figure}

Figure~\ref{fig:ion_h_radyn} shows the probability density function of the fraction of \HI, as function of temperature. There is a certain spread around the adopted fit which reflects the fact that the hydrogen ionzation is also a function of the history of the atmosphere. It is also clear that the coronal approximation gives a rather poor fit to the simulation data for temperatures below 30 kK.

\begin{figure}
 \includegraphics{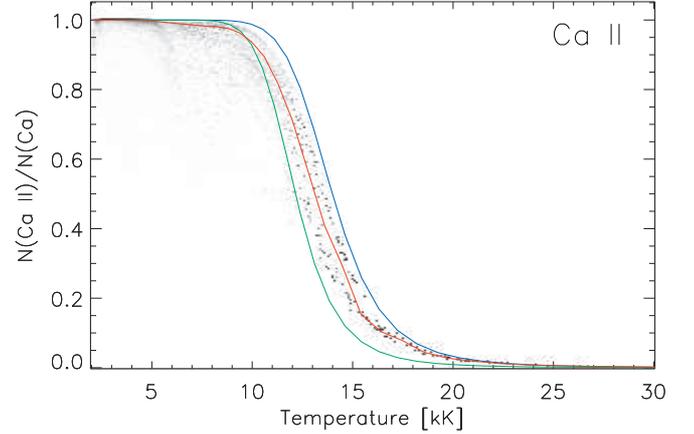}
 \caption{\label{fig:ion_ca} Probability density function of
   the fraction of Ca atoms in the form of \CaII\ as function of temperature in the \bifrost\ test
   atmosphere between heights of 0.5~Mm and 2.0~Mm. 
   Curves show the adopted fit to the PDF ({\it red}), the
   coronal approximation ({\it green}) and the coronal approximation with a
   two-level atom ({\it blue}). }
\end{figure}

Figure~\ref{fig:ion_ca} shows the same relation for \CaII. The points are rather close to the fitting function except for some points below 10 kK where an unusually strong radiation field may lead to a lower fraction of \CaII\ for some columns of the atmosphere. The best fit curve lies between the coronal approximation curve for a six-level atom (green curve) and the one for a two-level atom (blue curve). The reason the coronal approximation for the six-level atom fails is that an overpopulation of the meta-stable levels drives an upward radiative net rate in the infrared triplet violating the approximation of no upward radiative rates in the coronal approximation. In the limit where this upward rate is much larger than the downward rate we recover the two-level approximation.

\begin{figure}
 \includegraphics{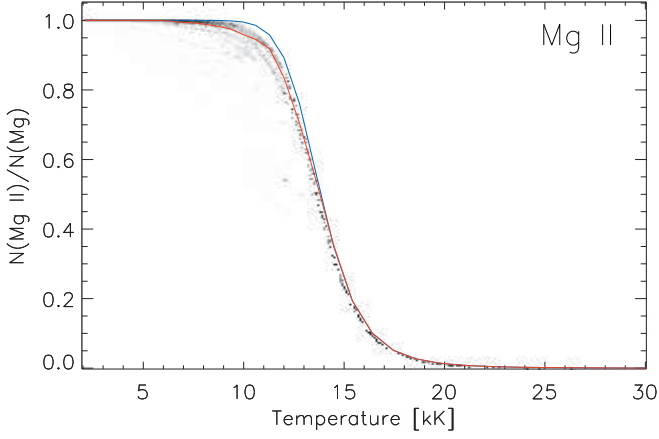}
 \caption{\label{fig:ion_mg}  Probability density function of
   the fraction of Mg atoms in the form of \MgII\ as function of temperature in the \bifrost\ test
   atmosphere between heights of 0.5~Mm and 2.0~Mm. 
   Curves show the adopted fit to the PDF ({\it red}), and 
   the coronal approximation ({\it blue}). }
\end{figure}

Figure~\ref{fig:ion_mg} shows the ionization balance of \MgII. The points are very close to the fitting curve over the full range of the simulation.

\section{Radiative heating}\label{sec:heating}

In the preceding section we deduced recipes to describe the radiative
cooling in transitions of \HI, \CaII\ and \MgII. 
In the cool phases of the chromosphere the atmosphere can be heated by
radiation in the same transitions.

The total radiative heating rate (radiative flux divergence) is given by
\be
Q=\int_0^\infty \chi_\nu(J_\nu-S_\nu)d\nu,\label{eq:qtot}
\ee
where $\chi_\nu$ is the opacity, $J_\nu$ is the mean intensity and $S_\nu$ is the
source function. Assuming a two-level atom and neglecting stimulated emission we have the following relations
\be
\chi_\nu=N_{{\rm X}_{m,i}}\alpha_0\phi_\nu=\frac{N_{{\rm X}_{m,i}}}{N_{{\rm X}_m}}
\alpha_0\phi_\nu\frac{N_{{\rm X}_m}}{N_{\rm X}}A_{\rm X}\frac{N_{\rm H}}{\rho}\rho
\ee
\be
S_\nu=\epsilon B_\nu + (1-\epsilon)J_\nu
\ee
\be
\epsilon=\frac{C_{ji}}{A_{ji}+C_{ji}}\approx\frac{C_{ji}}{A_{ji}}=f(T)n_e,
\ee
where $N_{{\rm X}_{m,i}}$ is the population density of the lower level $i$, $\alpha_0$ is
a constant, $\phi_\nu$ is the normalized absorption profile of the line, $\epsilon$ is
the collisional destruction probability, $B_\nu$ is the Planck function, $C_{ji}$ is the 
collisional deexcitation probability, $A_{ji}$ is the Einstein coefficient for spontaneous
emission, $f(T)$ is a function of temperature and the other terms have the same 
meaning as in Eq.~\ref{eq:Q}. Inserting into Eq.~\ref{eq:qtot} we get
\be
Q = f(T)\alpha_0 \frac{N_{{\rm X}_{m,i}}}{N_{{\rm X}_m}}
 \int_0^\infty\phi_\nu(J_\nu-B_\nu)d\nu 
 \frac{N_{{\rm X}_m}}{N_{\rm X}}A_{\rm X}\frac{N_{\rm H}}{\rho}n_e\rho.
\label{eq:qheat}
\ee
This equation resembles Eq.~\ref{eq:Q}; $\frac{N_{{\rm X}_{m,i}}}{N_{{\rm X}_m}}$ is
a function of temperature in LTE, $J_\nu$ is constant with height in the optically thin
regime such that $ \int_0^\infty\phi_\nu(J_\nu-B_\nu)d\nu$ is a function of temperature
(but also a function of the source function where $J_\nu$ thermalizes). At larger optical
depths, $J_\nu$ approaches $B_\nu$ and the integral behaves similarly to the
escape probability $E(\tau)$ in Eq.~\ref{eq:Q}. Although there are many approximations
involved, the similarity to Eq.~\ref{eq:Q} leads us to adopt the same functional form for
the heating as we have done for the cooling. We thus just use the previous fits also
when $L_{\rm X}(T)$ is negative (corresponding to heating) and adopt for simplicity
the same escape probability function.

\begin{figure}
 \includegraphics{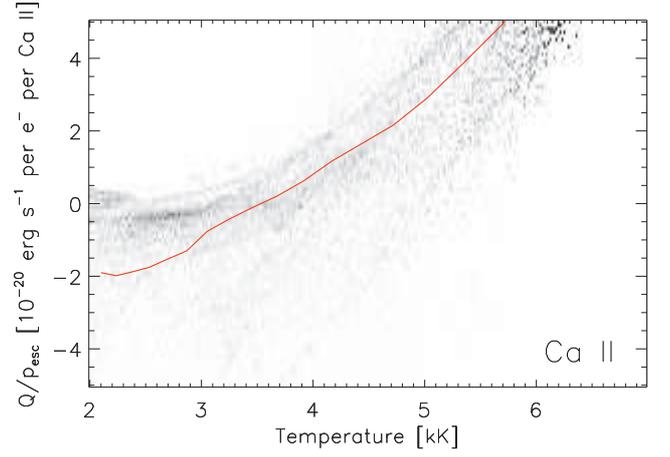}
 \caption{\label{fig:exc_ca_low_857} Probability density function
   of the radiative losses in lines of \CaII\ divided by the
   empirically determined escape probability as function of
   temperature in the \bifrost\ test atmosphere for points above
   0.3~Mm height. The red curve gives the adopted fit.}
\end{figure}

\begin{figure}
 \includegraphics{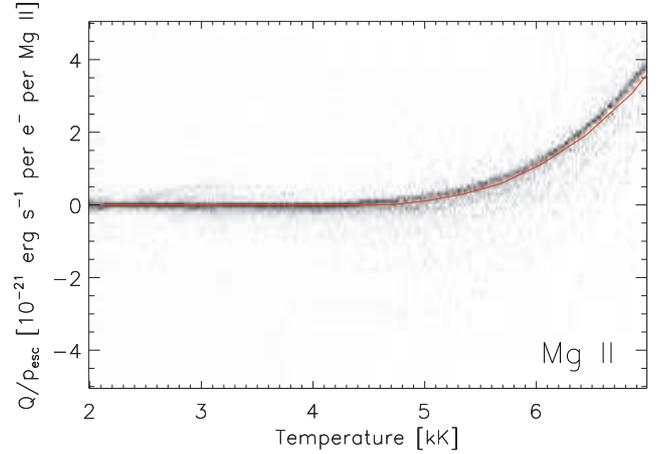}
 \caption{\label{fig:exc_mg_low_857} Probability density function
   of the radiative losses in lines of \MgII\ divided by the
   empirically determined escape probability as function of
   temperature in the \bifrost\ test atmosphere for points above
   0.3~Mm height. The red curve gives the adopted fit. }
\end{figure}

To test the applicability of this scheme we plotted the probability
density function of the radiative losses (positive) and gains
(negative) in the \bifrost\ test atmosphere, divided by the empirically
defined escape probability, as function of temperature. The results
are shown in Fig.~\ref{fig:exc_ca_low_857} for CaII\ and
Fig.~\ref{fig:exc_mg_low_857} for MgII. The figures show that there is
a fairly large spread in the losses/gains at low temperatures. For
\CaII\ we adopt a fit that gives gains below $ \sim 4$~kK, for
\MgII\ we choose zero gains/losses below $\sim 5$~kK.

Heating in hydrogen transitions is mainly due to absorption of
Lyman-$\alpha$ photons produced by a nearby source (a strong shock or
the transition region). This is illustrated in
Fig.~\ref{fig:cool_h_t=3170}, where there is heating due to
Ly-$\alpha$ photons caused by the emission at $^{10}\log m_\mathrm{c}
= -5.2$. Such behaviour cannot be modeled with a simple local equation
as Eq.~\ref{eq:qheat} so we choose to set the losses and gains for
hydrogen to zero below $\sim 4$~kK.

\section{Comparison between detailed solution and recipes}\label{sec:comparison}

We have now derived fits for the three factors that enter into the radiative cooling: the 
optically thin radiative loss function, the escape probability and the ionization fraction. 
To test the accuracy with which we can approximate the cooling calculated from the 
detailed solution of the non-LTE rate equations we here compare the detailed solution 
with our recipe. Since the detailed solution from a given test atmosphere has been used 
to derive the fits we here use different snapshots for this comparison (a different timeseries
calculated with RADYN for hydrogen and a snapshot at a much later time calculated with
\bifrost\ for calcium and magnesium).

\begin{figure}
 \includegraphics{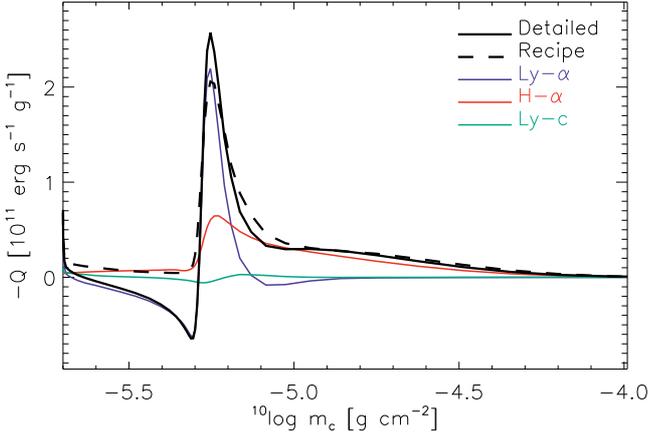}
 \caption{\label{fig:cool_h_t=3170}  Cooling as function of column mass from hydrogen 
 transitions at t=3170~s in the RADYN simulation. Total cooling from the detailed 
 calculation ({\it black solid}), from the recipes ({\it black dashed}), Lyman-$\alpha$
 ({\it blue}), H-$\alpha$ ({\it red}) and the Lyman continuum ({\it green}).}
\end{figure}

Figure~\ref{fig:cool_h_t=3170} shows the cooling ($-Q$) as a function of column mass for hydrogen
at a time in the RADYN simulation when there is a shock in the chromosphere. The detailed
cooling is dominated by Lyman-$\alpha$ in the shock but with a significant contribution also
from H-$\alpha$. Behind the shock (larger column mass), H-$\alpha$ dominates the cooling. 
The total cooling is well approximated by our recipe. In front of the shock (smaller column
mass) there is heating by absorption of Lyman-$\alpha$ photons. This heating is not 
accounted for in the recipe.

\begin{figure}
 \includegraphics{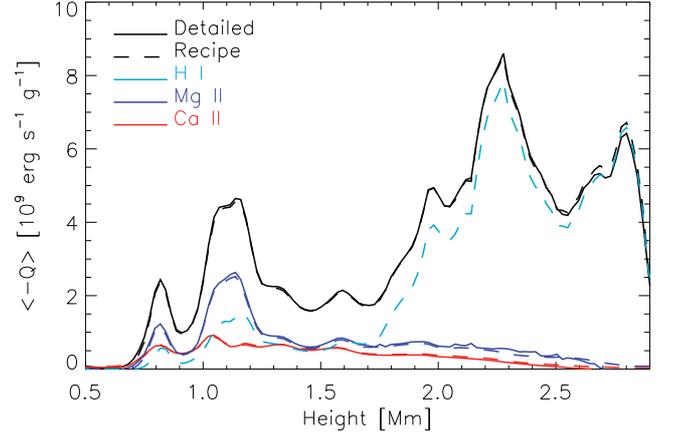}
 \caption{\label{fig:cool_z1}  Average cooling as function of height in the
 \bifrost\ test atmosphere. Total cooling  ({\it black}),  \HI\
 ({\it turquoise}), \MgII\ ({\it blue}) and \CaII \ ({\it red}). According to the detailed
 calculations ({\it solid}) and according to the recipes ({\it dashed}).}
\end{figure}

\begin{figure}
 \includegraphics{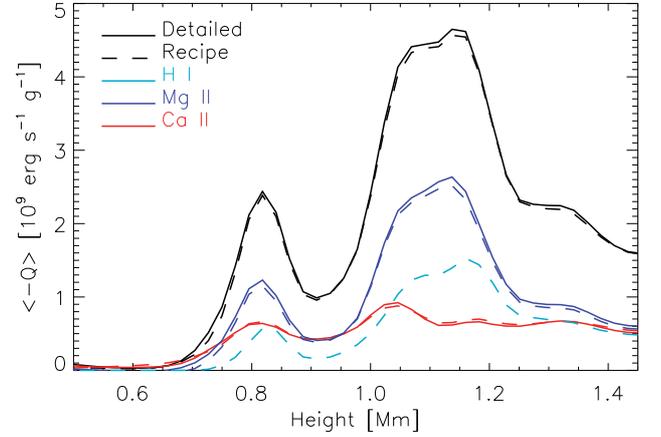}
 \caption{\label{fig:cool_z2}  As Fig.~\ref{fig:cool_z1} but for the
 low-mid chromosphere.}
\end{figure}

Figures~\ref{fig:cool_z1} and \ref{fig:cool_z2} show the average cooling as function of height
in the \bifrost\ test atmosphere. Since hydrogen cooling cannot be properly calculated in the
\bifrost\ atmosphere we only show the recipe result for hydrogen (and use these values also for the calculation of the total detailed cooling) in order to show the relative importance of the three elements. It is clear that the recipes reproduce the average cooling 
very well. Hydrogen dominates in the upper chromosphere and transition region while 
magnesium dominates the cooling in the mid and lower chromosphere with equally strong
calcium cooling at some heights. Note that this average cooling varies very much from snapshot 
to snapshot depending on the location of shocks in the atmosphere. These figures should
thus only be used to compare the detailed solution with the approximate one and not as 
pictures of the mean chromospheric energy balance.

\section{Heating from incident radiation field}\label{sec:incrad}

Part of the optically thin radiative losses from the corona escapes
outwards, but an equal amount of energy is emitted towards the sun
and is absorbed in the chromosphere where it contributes to radiative
heating. Most of this radiation is absorbed in the continua of neutral
helium and neutral hydrogen. If one assumes the coronal ionization
equilibrium is only dependent on temperature, then the
frequency-integrated coronal losses per volume
(positive Q means heating, as before) are given by
\be
Q_{\rm cor}= - n_{\rm H} \; n_{\rm e} f(T).
\ee
Here $f(T)$ is a positive function dependent on temperature only that can be
pre-computed and tabulated from atomic data. The emissivity associated
with this loss is 
\be
\eta = - \frac{Q_{\rm cor}}{4 \pi}.
\ee

Because the coronal radiative losses are integrated over frequency one
has to choose a single representative opacity to compute the
absorption. A decent choice is to use the
opacity at the ionization edge of the ground state of neutral helium
(denoted by $\alpha$). The opacity $\chi$ is then given by
\be
\chi=\alpha \, N_{\mathrm{He\, I}} = \alpha \, \rho \, {N_\mathrm{H} \over \rho}
A_{\mathrm{He}} {N_{\mathrm{He\,I}}\over N_{\mathrm{He}}},
\ee
where $N_{He I}$ is the number density of neutral helium, $N_{He}$ is
the number density of helium, $A_{He}$ is the abundance of helium
relative to hydrogen, $N_H$ is the total number density of hydrogen
particles and ${N_H\over \rho}$ is a constant dependent on the
abundances. The quantity $N_{\mathrm{He\,I}}/N_{\mathrm{He}}$ can be
pre-computed from a calibration computation as is done for \ion{H}{},
\ion{Ca}{} and \ion{Mg}{} in Sec.~\ref{sec:ionization}. 

The most accurate method for computing the absorption of
coronal radiation in the chromosphere is to compute the 3D radiation
field resulting from the coronal emissivity and the absorption using
the radiative transfer equation:
\be
\frac{\dd I}{\dd s} = \eta - \chi I
\ee
with $I$ the intensity and $s$ the geometrical path length along a
ray.  The solution can be computed using standard long or
short characteristics techniques in decomposed domains
\citep{Heinemann+Dobler+Nordlund+Brandenburg2006,Hayek+Asplund+Carlsson+Trampedach+etal2010}.
Once the intensity is known the heating rate can be
directly computed as
\be
Q_{\rm chr}  = \chi \int_\Omega I \dd \Omega - \eta = 4 \pi \, \chi \, J_\mathrm{3D} -\eta, \label{eq:chromheat}
\ee
with $\Omega$ the solid angle.

The method described above is accurate, but slow. A faster method
can be obtained by treating each column in the MHD simulation as a
plane-parallel atmosphere and assuming that at each location either
$\eta$ or $\chi$ is zero. For each column all quantities then become
only a function of height $z$ and Eq.~\ref{eq:chromheat} reduces to
\be
Q_{\rm chr}(z) = 2 \pi \, \chi(z) \int_{-1}^1
I(z,\mu) \dd \mu = 4 \pi \, \chi \, J_\mathrm{1D},
\label{eq:1dheat}
\ee
with $\mu$ the cosine of the angle with respect to the vertical. 

If we furthermore assume the emitting region is located on top of
the absorbing region, we can ignore rays that point outward
from the interior of the star in the integral of
Eq.~\ref{eq:1dheat}. The intensity that impinges on the top of the
absorbing region for inward pointing rays is simply the integral of
the emissivity along the $z$-axis from the top ($z_{\mathrm{t}}$) to
the bottom ($z_\mathrm{b}$) of the computational domain, weighted
with the inverse of $\mu$ to account for slanted rays
\be
I(\mu)_\mathrm{cor} = \frac{1}{|\mu|}\int_{z_{\mathrm{t}}}^{z_\mathrm{b}} \eta(z) 
\dd z \label{eq:corem}.
\ee

Now that we have computed the incoming intensity that impinges on the top 
of the absorbing region we can compute the
intensity at all heights from the solution to the transfer equation
with zero emissivity:
\be
 I(z,\mu) = I_\mathrm{cor}(\mu) \, \exp^{-\tau(z)/|\mu_i|}. \label{eq:chroabs}
\ee
The vertical optical depth $\tau$ is computed from the top of the
computational domain:
\be
\tau(z) = \int_{z_{\mathrm{t}}}^z \chi(z') \dd z'.\label{eq:tauvert}
\ee
The integrals in Eqs.~\ref{eq:corem} and~\ref{eq:tauvert} can both be
computed efficiently in parallel and need to be computed only once,
independent of the number of $\mu$-values in the angle discretization.

The method gives inaccurate results if the assumption of coronal
emission above chromospheric absorption fails. This happens for
example when a cool finger of chromospheric gas protrudes slanted
into the corona.

\begin{figure}
 \includegraphics{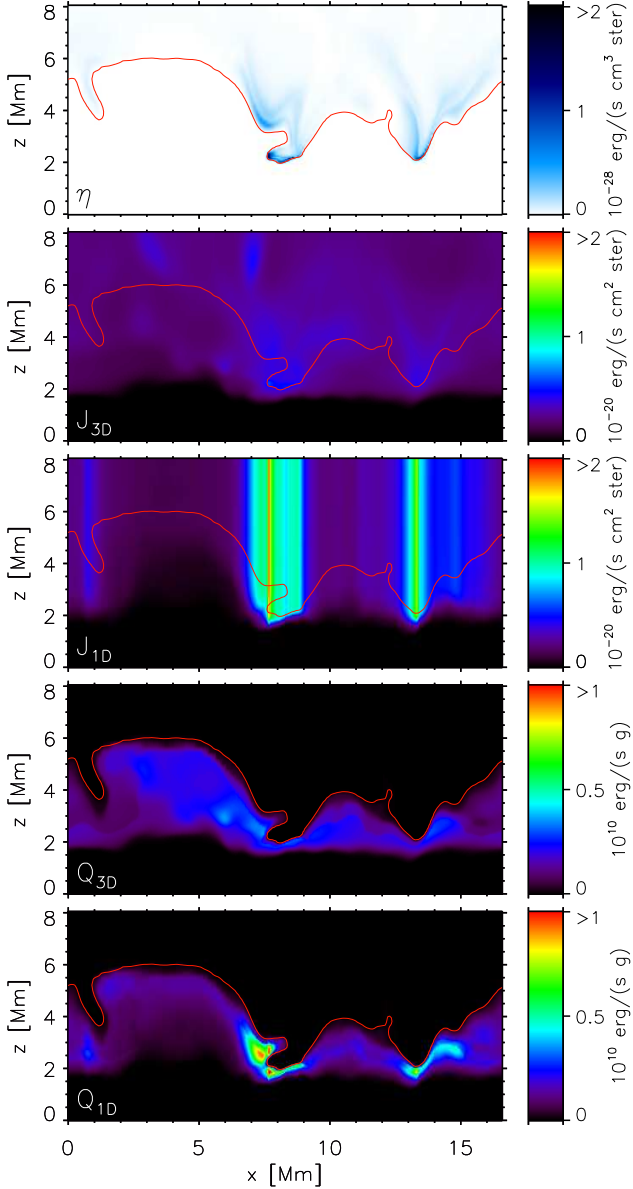}
 \caption{\label{fig:qincrad} Comparison of the 1D and 3D method for
   computing chromospheric heating due to absorption of coronal
   radiation. All panels show a vertical slice through a snapshot of a
   three-dimensional radiation-MHD simulation performed with
   \bifrost. The red curve is located at a gas temperature of 30,000~K
   and indicates the location of the transition zone.  All brightness
   ranges have been clipped to bring out variation at low values of the
   displayed quantities. Panels, from top to bottom: coronal
   emissivity $\eta$; angle-averaged radiation field computed in 3D;
   angle-averaged radiation field computed in 1D: heating rate per
   mass with 3D radiation: heating rate per mass with 1D
   radiation. Note that the 3D radiation field is computed using the
   full 3D volume, not only this 2D slice.}
\end{figure}

Figure~\ref{fig:qincrad} shows a comparison of the 3D and the
fast 1D method. The top panel shows the coronal emissivity $\eta$. It
is mostly smooth with occasional very large peaks close to the
transition region (red curve). The emissivity for gas colder than
30~kK (below the red curve) is negligible. 

The second panel shows the angle-averaged radiation field computed in
3D ($J_\mathrm{3D}$ in Eq.~\ref{eq:chromheat}). The radiation field is fairly smooth
as it is determined by an average of the emissivity throughout the
corona.

The middle panel shows the angle-averaged radiation field treating
each column as a plane-parallel atmosphere ($J_\mathrm{1D}$ in
Eq.~\ref{eq:1dheat}). Here it is far from smooth. The radiation field
is high in columns with a large integrated emissivity, as indicated by
the green and red vertical stripes. Note that there is cool gas
located above coronal gas around $(x,z)=(8,3)$~Mm. Here the radiation
field is highest at the top and decreases downward, while the
emissivity panel shows that the emission is mainly coming from the
coronal gas located below the cool finger.

The fourth panel shows the 3D heating rate. The heating rate is
non-zero only below the red curve, validating our assumption for the
fast 1D method of separate absorbing and emitting regions. It is
fairly smooth, with quite some heating in the cool bubble around
$(x,z)=(4,4)$~Mm, caused by coronal radiation coming in from the
sides.  The bottom panel shows the 1D heating rate. It shows large
extrema in columns with large integrated emissivity, and is less
smooth than the 3D heating. Because it is computed column-by-column, it cannot include the
sideways radiation in the cool bubble as indicated by the low heating
rate there compared to the 3D case. However, its overall appearance is
quite similar to the 3D heating rate.

\begin{figure}
 \includegraphics{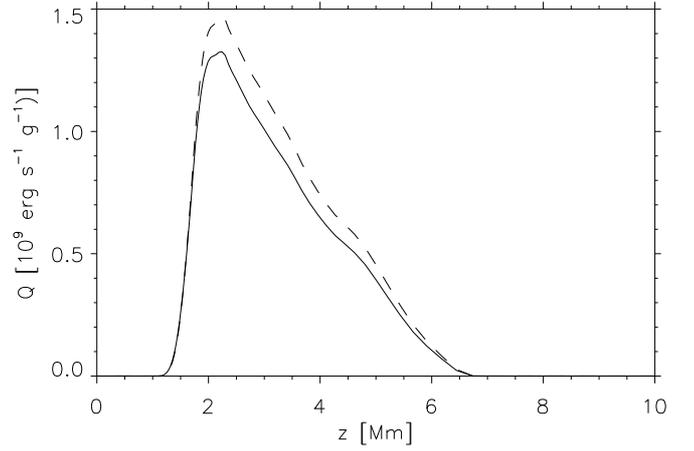}
 \caption{\label{fig:qincrad_avg} Horizontally-averaged radiative
   heating by absorption of coronal radiation as function of height in
   the 3D \bifrost\ test atmosphere from the detailed 3D computation ({\it solid}) and
   from the 1D recipe ({\it dashed}).}
\end{figure}

Figure~\ref{fig:qincrad_avg} compares the horizontally
averaged heating rate as function of height computed from the 1D
recipe with the 3D heating rate. The recipe is always within 15\% of
the 3D heating. Part of this difference is caused by numerical errors
because the strong coronal emission close to the transition region is
only barely resolved on the grid used in the simulation.

\section{Conclusions} \label{sec:conclusions}

We have shown that the radiative cooling in the solar chromosphere 
by hydrogen lines, the
Lyman continuum, the H and K and infrared triplet lines of singly
ionized calcium, and the h and k lines of singly ionized magnesium
can be expressed fairly accurately as easily computed
functions. These functions are the product of an optically thin emission
(dependent on temperature), an escape probability (dependent on column
mass) and an ionization fraction (dependent on temperature). While at
any given location in the atmosphere our recipe might deviate from the
true cooling, it is nevertheless remarkably accurate when averaged
over time and/or space as indicated by Figs.~\ref{fig:cool_h_t=3170}
-- \ref{fig:cool_z2}.

We have shown that radiative heating can be approximated with the same
functional form as radiative cooling. However, the heating is strongly
dependent on the details of the local radiation field and absorption
profile, so the recipe is not nearly as accurate for heating as it is
for cooling. Our detailed computations show that radiative heating in
the h and k lines of \MgII\ is negligible, and small, but present, for
the lines of \CaII. Heating due to hydrogen is mainly through
absorption of  Lyman-$\alpha$ photons next to the transition region or
shock waves. This effect cannot be modeled with our recipes.

In addition we have presented a simple method to compute chromospheric
heating by absorption of coronal radiation. This method approximates
the diffuse component of the heating, but yields localized spots of
spurious intense heating owing to the column-by-column approach that
prevents spreading of radiation. It reproduces the
spatially averaged heating very well. Given its simplicity and speed it is
a good alternative to the slower, more accurate method that requires a
formal solution of the transfer equation.

The recipes have been developed for solar chromospheric conditions. For 
sunspot conditions or for other stars it would be necessary to redo the 
semi-empirical fits using detailed radiative transfer calculations applied to 
appropriate simulations.
 
\begin{acknowledgements}
This research was supported by the Research Council of Norway through
the grant ``Solar Atmospheric Modelling'' and 
through grants of computing time from the Programme for Supercomputing.
J.L. acknowledges financial support from the European Commission
through the SOLAIRE Network (MTRN-CT-2006-035484) and  from the Netherlands Organization for
  Scientific Research (NWO).
\end{acknowledgements}

\bibliographystyle{aa}

\end{document}